# Rapid Reconstruction of Extremely Accelerated Liver 4D MRI via Chained Iterative Refinement

Di Xu[1], Xin Miao[2], Hengjie Liu[3], Jessica E. Scholey[1], Wensha Yang[1], Mary Feng[1], Michael Ohliger[1], Hui Lin[1], Yi Lao[3], Yang Yang[4] and Ke Sheng[1, *]

1 Radiation Oncology, University of California, San Francisco, 505 Parnassus Ave, San Francisco, CA 94143

2 Siemens Healthineers, 1 Federal St, Boston, MA 02110

3 Radiation Oncology, University of California, Los Angeles, 200 Medical Plaza, Los Angeles, CA 90095

4 Radiology, University of California, San Francisco, 505 Parnassus Ave, San Francisco, CA 94143

Corresponding author: ke.sheng@ucsf.edu

# Abstract

**Purpose**: 4D MRI with high spatiotemporal resolution is vital to characterize the tumor/tumor motion for liver radiotherapy. However, high-quality 4D MRI requires an impractically long scanning time for dense k-space signal acquisition covering all respiratory phases. Accelerated sparse sampling followed by reconstruction enhancement is desired but often results in degraded image quality and long reconstruction time. We hereby propose the chained iterative reconstruction network (CIRNet) for efficient sparse-sampling reconstruction while maintaining clinically deployable quality.

**Methods**: CIRNet adopts the denoising diffusion probabilistic framework to condition the image reconstruction through a stochastic iterative denoising process. During training, a forward Markovian diffusion process is designed to gradually add Gaussian noise to the densely sampled ground truth (GT), while CIRNet is optimized to iteratively reverse the Markovian process from the forward outputs. At the inference stage, CIRNet performs the reverse process solely to recover signals from noise, conditioned upon the undersampled input. CIRNet is structured with a U-Net architecture, optimized to minimize the $L_2$ difference between estimated and GT noises. CIRNet processed the 4D data (3D+t) as temporal slices (2D+t). The proposed framework is evaluated on a data cohort consisting of 48 patients (12332 temporal slices) who underwent free-breathing liver 4D MRI. 3-, 6-, 10-, 20- and 30-times acceleration were examined with a retrospective random undersampling scheme. Compressed sensing (CS) reconstruction with a spatiotemporal constraint and a recently proposed deep network, Re-Con-GAN, are selected as baselines.

**Results**: CIRNet consistently achieved superior performance compared to CS and Re-Con-GAN (e.g., PNSR of CIRNet, CS and Re-Con-GAN is at 22.35±2.94, 13.27±3.89 and 13.27±3.89 dB in 30 times acceleration). The inference time of CIRNet, CS, and Re-Con-GAN are 11s, 120s, and 0.15s.

**Conclusion**: A novel framework, CIRNet, operating under stochastic iterative refinement for accelerated MR reconstruction, is presented. Compared with published methods, CIRNet maintains useable image quality for acceleration up to 30 times, significantly reducing the burden of 4DMRI.

1. Introduction

Owing to its superior soft tissue contrast compared to computed tomography, magnetic resonance imaging (MRI) has been increasingly adopted for image-guided liver radiation therapy (RT)[1–3]. 4D MRI, a volumetric imaging technique for respiratory-resolved images, is suited for characterizing tumor morphologies and motions[4–6]. During the planning for free-breathing liver RT, the internal target volume (ITV) for free-breathing or gated treatments is determined based on individually contoured 4D MRIs. An inaccurately defined ITV may lead to insufficient tumor coverage or over-dosing of the surrounding normal tissue.

A typical motion-insensitive technique for 4D MRI acquisition is via continuous free-breathing scan of 3D golden angle stack-of-stars spokes, where radial sampling is conducted in the $kx - ky$ plane to reduce motion sensitivity[7] and incoherent k-space understanding[8]; Cartesian sampling is applied in the $kz$ plane to allow for flexible selection of volumetric coverage[9]. Yet, such scans usually take 8-10 minutes to achieve acceptable spatiotemporal resolution for all respiratory phases[5]. The long scan time is a significant burden to patients, considering the need to acquire other essential images. Though undersampling in the $kx - ky$ plane can effectively reduce the scan time, it also leads to streaking artifacts, uneven brightness, and blurriness using conventional reconstruction methods. Parallel imaging[10,11], which simultaneously acquires numerous views with multiple receiver coils, and compressed sensing (CS)[12–15], a constrained iterative optimization framework, have been previously proposed to mitigate these problems. Still, the images are unusable with high acceleration ratios[16].

In the past decade, numerous deep learning (DL) based algorithms have been proposed for 4D MRI reconstruction. Ample studies have demonstrated that well-trained DL models can match/exceed CS performance with significantly faster reconstruction[16–18]. Existing works have explored using convolutional neural networks (CNNs)[18], recurrent neural networks (RNNs)[16], Transformers[19], or generative adversarial training (GAN) assisted networks for 4D MR reconstruction[20]. For instance, Schlemper et al. proposed a cascade 3D CNN architecture for 4D MRI reconstruction[18] and demonstrated their model performance in a dataset with up to 11-fold undersampling. Adapting from the UNet architecture, Dracula[21] and Moivenet[22] were proposed to accelerate 4D MR reconstruction using multi-coil images with 1.5, 2 and 2.5-fold acceleration explored. Moreover, Huang et al. introduced a motion-guided framework using RNN-inspired Conv-GRU for initial 2D frame reconstruction and U-FlowNet for motion estimation in the optical flow field[23]. The proposed pipeline reconstructed a 5 and 8 times accelerated (5x and 8x) cardiac dataset. Xu et al. then proposed a 2D CNN-assisted Reconstruction Swin Transformers (RST), a variant of Video Swin Transformers[24], and validated the algorithm on a 9x accelerated cardiac dataset[19]. Xu et al. recently explored a GAN-based framework (Re-Con-GAN) with up to 10x acceleration on liver 4D MRI pursued[20]. However, the performance of most existing methods declines rapidly beyond 10x. Since those frameworks require carefully designed regularization and optimization tricks, which are often hard to seek in challenging tasks to tame the optimization instability[25,26] and avoid model collapsing[27,28].

Recently, stochastic diffusion probabilistic models, such as Denoising Diffusion Probabilistic Model (DDPM)[29,30] and Super Resolution via Repeated Refinement (SR3)[31], have been demonstrated to be superior in natural image super-resolution tasks compared to the regression-based CNNs/RNNs/Transformers and GAN enhanced frameworks. Such diffusion architectures have two major advantages. 1) Modelling imaging noise instead of the mean to form a better-defined optimization target. 2) Incrementally approaching the optimization target for better modeling stability. A few pioneer works have explored the possibility of using diffusion framework for MRI reconstruction[32–35]. In specific, Chung et al. proposed a score-based diffusion model for score function assisted diffusion in accelerated 3D MRI reconstruction[32]. Gungor et al. proposed AdaDiff, a two-phrase domain-shift resistant framework, for 3D multi-contrast brain MRI reconstruction[33]. Korkmaz et al. proposed SSDiffRecon, a self-supervised diffusion-based framework for reconstruction without supervision from fully sampled sequences. SSDiffRecon demonstrates its performance on 3D brain MRI dataset[34]. Additionally, Zhao et al. proposed DiffGAN, a GAN based diffusion model for stabilized GAN based MRI training. The proposed framework was evaluated on 3D MRI reconstruction of various anatomies[35]. Though fruitful results achieved, neither of the previous studies push the acceleration ratio beyond 10x nor do they apply the diffusion mechanism in the setting of 4D MRI reconstruction.

To this end, we hereby propose a chained iterative refinement network (CIRNet), a framework rooted in stochastic diffusion, to achieve the reconstruction in ultra-sparse undersampled 4D MRI sequences. The proposed framework is designed to learn the two-dimensional temporal image series (2D+t) from under-sampled data. Experiments on an in-house 4D liver MRI dataset demonstrate the superior performance of CINet compared to conventional CS and a state-of-the-art DL reconstruction model (Re-Con-GAN[20]). The rest of the manuscript is organized as follows: Section 2 elaborates on data cohort, CIRNet framework, baseline algorithms, and model evaluation; Section 3 summarizes the experimental results; and Section 4, along with Section 5, discusses and concludes the current work.

## 2. Materials and Methods

### 2.1 Data Cohort

The study was approved by the local Institutional Review Board at UCSF (#14-15452). 48 patients were scanned on a 3T MRI scanner (MAGNETOM Vida, Siemens Healthcare, Erlangen, Germany) after injecting hepatobiliary contrast (gadoxetic acid; Eovist, Bayer). A prototypical free-breathing T1-weighted volumetric golden angle stack-of-stars sequence was used for 4D MRI acquisition. The scanning parameters were - TE=1.5 ms, TR=3 ms, matrix size = 288x288, FOV = 374 mm x 374 mm, in-plane resolution=1.3 mm × 1.3 mm, slice thickness=3 mm, radial views (RV) per partition=3000, number of slices or partitions = 64-75, acquisition time = 8-10 min. The pulse sequence ran continuously over multiple respiratory cycles. Images reconstructed from the entire space data of 3000 radial spokes (RV-3000) were treated as the fully sampled ground truth reference (Based on Nyquist sampling theorem, fully sampled radial images require sampling points $\times \frac{\pi}{2}$ spokes, resulting in 452 spokes for a matrix size of $288 \times 288$. After motion binning, each of the eight bins has, on average, 375 spokes with RV3000, close to fully sampled 452 spokes). Retrospective under-sampling was performed by randomly selecting 1000, 500, 300, 150, and 100 spokes from the 3000 spokes, corresponding to acceleration ratios of 3x, 6x, 10x, 20x, and 30x. For initial image reconstruction, data sorting based on a self-gating signal was performed to divide the continuously acquired k-space data into 8 respiratory phases. nonuniform fast Fourier transform (nuFFT) algorithm was applied to reconstruct each phase individually. Only regular breathers (48 patients) were included in the current project. Breathing regularity was quantified using the self-gating signal waveform[36–38]. The peak-to-trough range and mid-level amplitude (A), i.e., (peak-A + trough-A)/2, were calculated for each respiratory cycle. The average mid-level amplitude across all respiratory cycles normalized with the average peak-to-trough range was used as the regularity measurement. Patients with a score greater than 20% were classified as irregular breathers and excluded.

To augment the sample size, we organized the 48 3D+t data as 12332 2D+t images with images from an individual patient sorted in one subset. The data was split into training (37 patients with 10721 2D+t images) and testing (11 patients with 1611 2D+t images), where patients with various profiles (body mass index and breathing regularity score) are balanced in each split. The images were resized to $256 \times 256$ and normalized using Z-score normalization.

## 2.2 CIRNet Framework

The source-target image sequence pairs, $D = \{x_i, y_i\}_{i=1}^{N}$, are samples drawn from the unknown distribution $p(y|x)$. Our goal is to learn a parametric approximation of $p(y|x)$ that can map $x$ to $y$. We approach this problem through the forward and backward stochastic iterative diffusion refinement process. As seen in Figure 1, the forward diffusion process $q$ incrementally adds Gaussian noise to the fully sampled image sequence $y_0$ over $T$ timesteps through a fixed Markov chain $p_\theta(y_t|y_{t-1})$. The backward diffusion process $p_\theta$ is parameterized with CIRNet, which aims to iteratively recover signals from noise $y_T$ conditioned on the information in source $x$.

The CIRNet architecture is a modified U-Net from SR3[31]. We replace the original residual blocks with that from BigGAN[39], rescale the skip connections by $\frac{1}{\sqrt{2}}$, and replace the input/output number of channels ($C = 3$) in natural images with the number of motion bins ($C = T = 8$) in our image sequence. Therefore, the input to the network is of shape $C(T) \times H \times W \times Z = 8 \times 256 \times 256 \times 1$, where $H$, $W$ and $Z$ are the input height, width and number of slices, respectively. The forward training noise schedule uses a piece-wise distribution $\gamma$ that conducts uniformly sampling through the iteration timesteps $T$, where $T = 800$ across our experiments.

Assuming the forward diffusion process can be viewed as a fixed approximate posterior to the inference process, we derive the variational lower bound on the marginal log-likelihood as Equation (1). Given a particular parameterization $p_\theta$, we derive the negative variational lower bound as the simplified $L_2$ loss for $p_\theta$ optimization, defined in Equation (2), up to a constant weighting of each term for each time step[29].

$$E_{(x,y_0)} \log p_\theta(y_0|x) \geq E_{x,y_0} E_{q(y_{1:T}|y_0)} \left[ p_\theta(y_T) + \sum_{t \geq 1} \log \frac{p_\theta(y_{t-1}| y_t, x)}{q(y_t|y_{t-1})} \right] \quad (1)$$

$$E_{x,y_0,\epsilon} \sum_{t=1}^{T} \frac{1}{T} ||\epsilon - \epsilon_\theta(x, \sqrt{\gamma_t} y_0 + \sqrt{1 - \gamma_t} \epsilon, \gamma_t)||_2^2 \quad (2)$$

Where $\epsilon \sim \mathcal{N}(0, \boldsymbol{I})$.

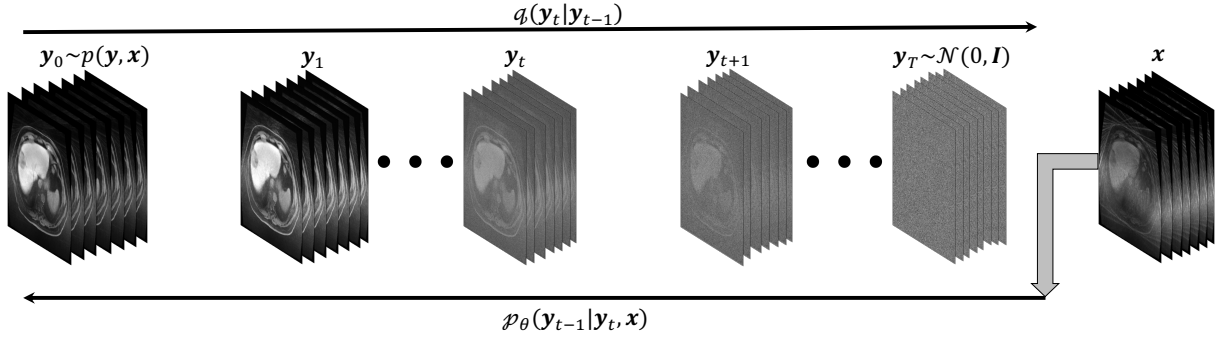

**Figure 1**: CIRNet structure. The forward diffusion process $q$ (left to right) gradually adds Gaussian noise $\mathcal{N}(0, \boldsymbol{I})$ to the target fully sampled image sequence $\boldsymbol{y_0}$ over $T$ timesteps. The reverse inference process $p_\theta$ (right to left) iteratively denoises the noisy target images $\boldsymbol{y_T}$ conditioned on the source undersampled image sequence $\boldsymbol{x}$.

## 2.3 Model Training

CIRNet was trained for 1,000,000 iterations with a batch size of $4 \times 1$ across all the acceleration ratios, with the best performer selected via cross-validation saved as final model weights. Adam optimizer was used with a linear warmup schedule of over 10,000 training iterations, followed by a fixed learning rate of 0.0001 for the rest of the training. The model was implemented in PyTorch with all the experiments carried out on a $4\times RTXA6000$ GPU cluster.

## 2.4 Baseline Algorithms and Model Evaluation

A conventional CS algorithm and a GAN-based DL method, Re-Con-GAN[20], were included as benchmarks. For Re-Con-GAN, ResNet9 generator was selected. The CS algorithm was implemented with close-sourced Siemens ICE platform, while Re-Con-GAN was implemented with the author's original experimental codes.

The model performance was evaluated using the following metrics: root mean squared error (RMSE), peak-signal-to-noise ratio (PSNR), structure similarity indexed measurement (SSIM), and inference time, as shown in Equation (3-5).

$$RMSE = \sqrt{\frac{\sum_{i=1}^{N}(G(x,z) - y)^2}{N}} \qquad (3)$$

$$PSNR = 20 \cdot \log_{10} \frac{MAX_I}{RMSE} \qquad (4)$$

$$SSIM = \frac{(2\mu_{G(x,z)}\mu_y + c_1)(2\sigma_{G(x,z)y} + c_2)}{(\mu_{G(x,z)}^2 + \mu_y^2 + c_1)(\sigma_{G(x,z)}^2 + \sigma_y^2 + c_2)} \quad (5)$$

Where $MAX_I$ is the maximum possible pixel value in a tensor, $\mu_{G(x,z)}$ and $\mu_y$ is the pixel mean of $G(x,z)$ and $y$ and $\sigma_{G(x,z)y}$ is the covariance between $G(x,z)$ and $y$, $\sigma_{G(x,z)}^2$ and $\sigma_y^2$ is the variance of $G(x,z)$ and $y$. Lastly, $c_1 = (k_1 L)^2$ and $c_2 = (k_2 L)^2$, where $k_1 = 0.01$ and $k_2 = 0.03$ in the current work, and $L$ is the dynamic range of the pixel values ($2^{\# \, bits \, per \, pixel} - 1$).

## 3. Results

The quantitative results and selected visualization of the test set are reported in Table 1, Figure 2, and Figure 3. Visually, Figure 2 shows that as the acceleration ratio increases from 3x to 30x, the under-sampled nuFFT images (first column) become unusable with rapidly increasing streaking artifacts and image noises. In comparison, CIRNet is robust to increasing undersampling, showing useable images at the highest acceleration with a marginally decreased detail retention and slightly more noticeable artifacts and noises. CIRNet consistently outperforms comparison methods both visually and quantitatively. The Re-Con-GAN has smoother predictions with notable detail loss at higher acceleration ratios (10x and 30x). At 3x acceleration, CIRNet achieves superior 1-SSIM (0.04) and RMSE (0.06), with well-preserved anatomical details. As acceleration increases, CIRNet maintains better quantitative performance (e.g., 1-SSIM: 0.05 at 6x, 0.06 at 10x, 0.08 at 20x and 0.11 at 30x), and image detail retention, compared with more pronounced detail loss and degraded quantitative performance by Re-Con-GAN (1-SSIM: 0.07 at 6x, 0.12 at 10x, 0.16 at 20x and 0.17 at 30x) and CS (1-SSIM: 0.08 at 6x, 0.13 at 10x, 0.18 at 20x and 0.19 at 30x). Speed-wise, the averaged reconstruction time of CIRNet for a 4D volume (around 11 s) is around 10 times faster than CS (120 s), though less efficient than the Re-Con-GAN (sub-second), which does not require test-phase optimization. Figure 3 shows that CIRNet is robust across all acceleration levels and can maintain useable contrast and sharpness across all motion phases.

| Model | Acceleration | PSNR (dB) ↑ | 1-SSIM ↓ | RMSE ↓ | Inference Time (s) ↓ |
|---|---|---|---|---|---|
| CS | 3x | 25.31±2.56 | 0.07±0.02 | 0.17±0.05 | 120 |
| Re-Con-GAN | | 26.13±3.02 | 0.05±0.02 | 0.08±0.03 | **0.15** |
| CIRNet | | **29.43±2.67** | **0.04±0.02** | **0.06±0.02** | 11 |
| CS | 6x | 20.73±2.95 | 0.08±0.02 | 0.19±0.05 | |
| Re-Con-GAN | | 23.97±3.84 | 0.07±0.03 | 0.16±0.05 | |
| CIRNet | | **28.36±2.78** | **0.05±0.02** | **0.07±0.03** | |
| CS | 10x | 19.29±2.99 | 0.13±0.05 | 0.21±0.08 | - |
| Re-Con-GAN | | 21.61±2.93 | 0.12±0.03 | 0.13±0.04 | |
| CIRNet | | **26.36±2.89** | **0.06±0.03** | **0.09±0.03** | |
| CS | 20x | 15.35±3.67 | 0.18±0.09 | 0.26±0.15 | |
| Re-Con-GAN | | 17.02±3.45 | 0.16±0.07 | 0.22±0.12 | |
| CIRNet | | **25.03±2.92** | **0.08±0.05** | **0.12±0.05** | |
| CS | | <u>13.27±3.89</u> | <u>0.19±0.12</u> | <u>0.29±0.19</u> | |
| Re-Con-GAN | 30x | 15.89±3.65 | 0.17±0.09 | 0.25±0.15 | |
| CIRNet | | **22.35±2.94** | **0.11±0.07** | **0.15±0.07** | |

**Table 1**: Statistical results from our proposed CIRNet under 3x, 6x, 10x, 20x and 30x acceleration rate and their corresponding baselines are presented. The best score and the worst score under each acceleration is bolded and wavy underlined, respectively. The up arrows next the evaluation metrics means that a higher value is superior and vice-versa for the down arrow. All the statistics are calculated with images normalized to [0,1] scale.

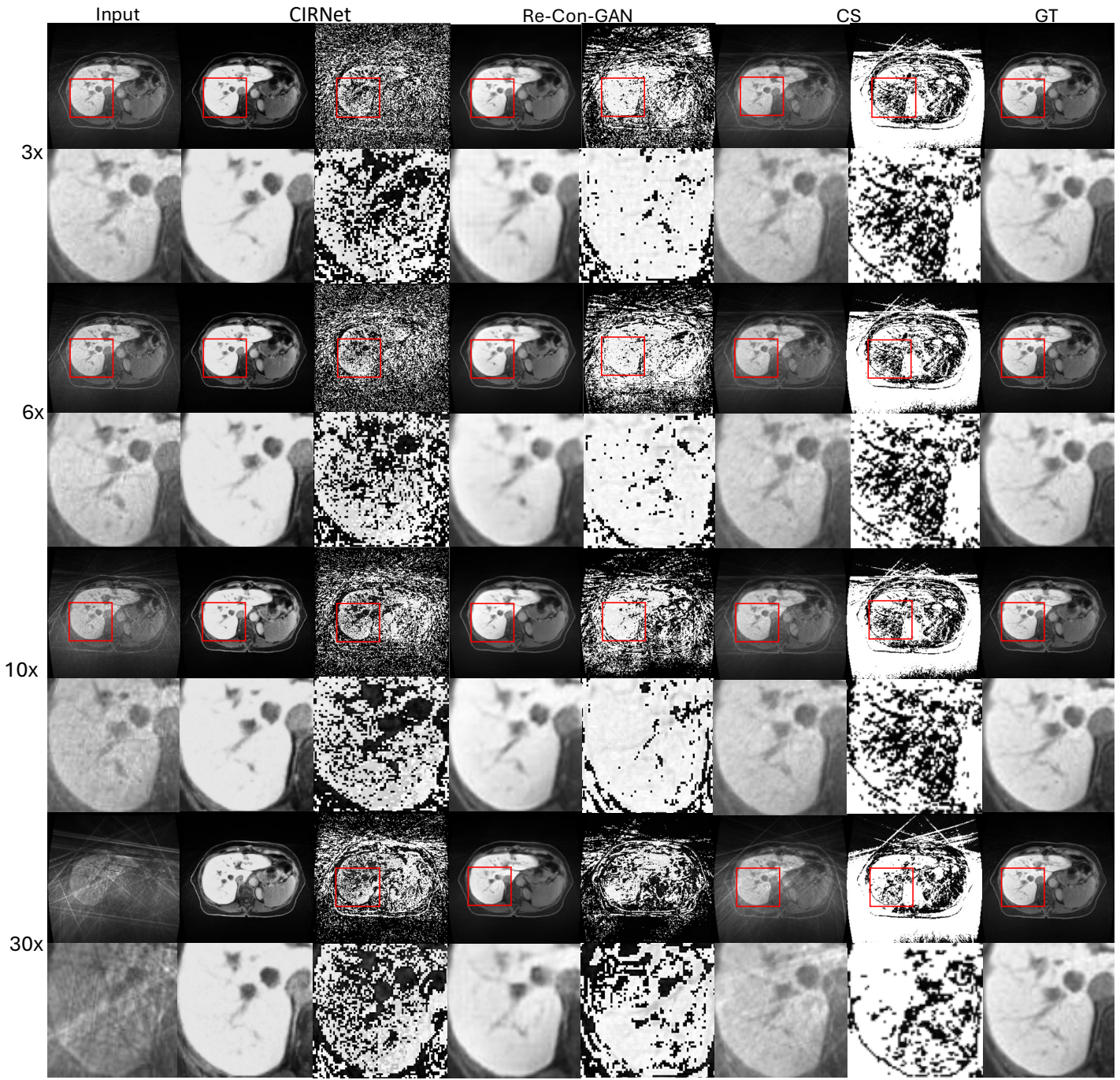

**Figure 2**: Visualization of 3x, 6x, 10x and 30x reconstruction results from CIRNet, Re-Con-GAN and CS of an axial view slice from a patient in the test set. Reconstruction visualization, zoomed-in region of interest (regions in the red boxes) as well as residual between prediction and the fully sampled GT are visualized. All the images are visualized after normalizing to [0,1] scale.

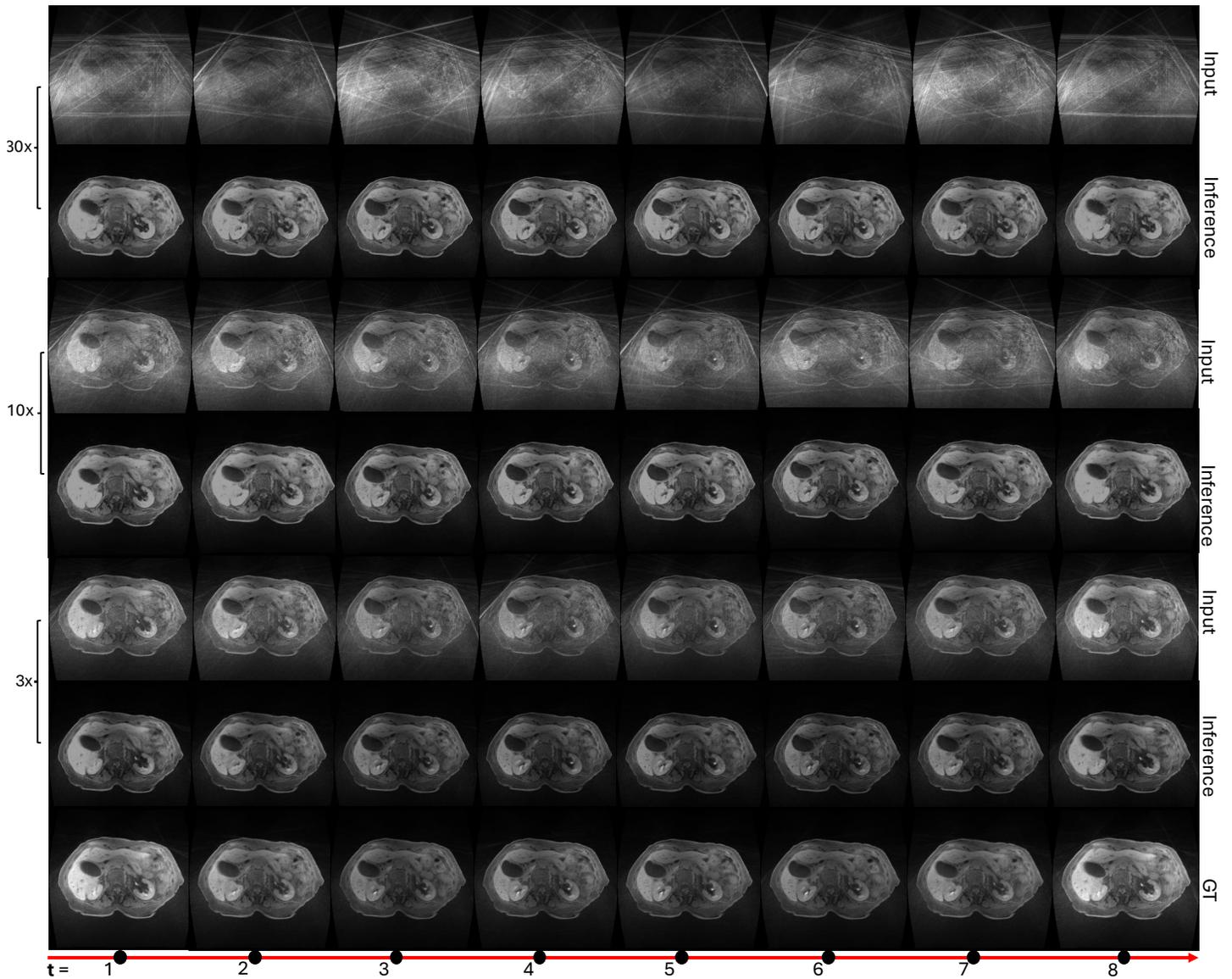

**Figure 3**: Visualization of the temporal profile of a patient in the test set. 3x, 10x, and 30x reconstruction results from input, GT, and our proposed CIRNet are visualized. All the images are visualized after normalizing to [0,1] scale.

## 4. Discussion

Hepatocellular carcinoma (HCC) ranks at the top ten most prevalent cancers globally, and it has become one of the leading and fastest-growing causes of cancer-related death[40,41]. Moreover, the liver is a frequent site for metastases from various cancer types, including colorectal, pancreatic, melanoma, lung, and breast cancers[42]. Although surgical resection remains the standard of care for HCC[43], RT is an effective alternative for unresectable patients[44]. Owing to its superior soft-tissue contrast, 4D MRI is a crucial imaging tool for image-guided RT (IGRT) in liver cancer.

Though existing DL and analytical methods reported promising results for accelerated 4D MRI reconstruction, none maintain usable image quality with >10x acceleration. The current paper shows that CIRNet can maintain its reconstructed image quality up to 30x acceleration, leading to around 20-second acquisition, significantly reducing the patient burden. There are several theoretical and practical advantages to using CIRNet for 4D MRI reconstruction. First, CIRNet demonstrates superior retention of subtle tissue textures, which is essential for the accurate delineation of liver tumors. In comparison, Re-Con-GAN suffers from detail loss due to the inherent smoothing effects of CNNs, while CS methods tend to lose the structures and leave more pronounced artifacts that worsen steeply with high acceleration rates. Second, CIRNet offers significantly faster reconstruction (11s) than the CS method (120s), making them practical for time-sensitive adaptive IGRT. Though CIRNet is slower than Re-Con-GAN (<1s) inference-wise, CIRNet offsets longer reconstruction time with shorter acquisition. The acquisition time using CIRNet approaches the duration of a breathing cycle, indicating the potential of capturing real-time motion with minimal binning/averaging.

Nevertheless, the current work is not without room for improvement. First, our implementation is limited to learning 2D+t image series. Training with 3D+t data would enable more effective learning of inter-slice anatomy but would demand an impractically large GPU memory capacity. Second, the current method functions in the image domain, which leads to potential information loss and added latency due to preliminary k-space to image transformation. Future work will explore frameworks working in k-space or using multi-coil data as the model input. Lastly, CIRNet can be complex and less interpretable, especially when trained on 4D data, which could pose potential challenges in clinical validation, where model interpretability is essential. In the future, explainable artificial intelligence techniques, such as gradient-weighted class activation mapping[45], saliency maps[46], and feature attribution analysis[47], can be incorporated to improve the transparency and trustworthiness of the current framework.

## 5. Conclusion

An ultra-sparse liver 4D MRI reconstruction framework, CIRNet, is proposed in the current work. CIRNet employs a stochastic diffusion process to iteratively model the source noise for robust image reconstruction. The evaluation conducted on an in-house data cohort demonstrates promising imaging quality at ultra-high acceleration ratios.